\documentclass[conference]{IEEEtran}
\usepackage[utf8]{inputenc}
\usepackage{amsmath}
\usepackage{graphicx}
\usepackage{algorithm}
\usepackage{algpseudocode}
\usepackage{hyperref} % for email
\usepackage{cite}
\usepackage{booktabs}
\usepackage{multicol}
\usepackage{multirow}
\usepackage{array}
\usepackage{cellspace}
\usepackage{eso-pic}
\newcommand\AtPageUpperMyleft[1]{\AtPageUpperLeft{
 \put(\LenToUnit{1cm},\LenToUnit{-1cm}){ % Adjust 1cm for fine-tuning
     \parbox{0.5\textwidth}{\raggedright\fontsize{9}{11}\selectfont #1}} % Changed from \raggedleft to \raggedright
 }}
\newcommand{\conf}[1]{
\AddToShipoutPictureBG*{
\AtPageUpperMyleft{#1}
}
}

\begin{document}

\title{Benchmarking ChatGPT, Codeium, and GitHub Copilot: A Comparative Study of AI-Driven Programming and Debugging Assistants}
% \conf{This work has been submitted to the IEEE for possible publication. Copyright may be transferred without notice, after which this version may no longer be accessible.} 
\conf{This work has been submitted to the IEEE for possible publication. Copyright may be transferred without notice, after which this version may no longer be accessible.} 

\author{\IEEEauthorblockN{
Md Sultanul Islam Ovi\IEEEauthorrefmark{1},
Nafisa Anjum\IEEEauthorrefmark{3},
Tasmina Haque Bithe\IEEEauthorrefmark{2},\\
Md. Mahabubur Rahman\IEEEauthorrefmark{2}, and  
Mst. Shahnaj Akter Smrity\IEEEauthorrefmark{2}
}
\IEEEauthorblockA{\IEEEauthorrefmark{1}Dept. of Computer Science, George Mason University, Fairfax, Virginia, USA}
\IEEEauthorblockA{\IEEEauthorrefmark{3}Dept. of Electrical \& Electronic Engineering, Rajshahi University of Engineering \& Technology, Bangladesh}
\IEEEauthorblockA{\IEEEauthorrefmark{2}Dept. of Computer Science and Engineering, Green University of Bangladesh, Bangladesh}
\IEEEauthorblockA{Email: movi@gmu.edu, nafisaanjum9999@gmail.com, tasmina.haque768@gmail.com,\\ mahabuburrahman0002@gmail.com, kazishahnajs@gmail.com}
}

\maketitle

\begin{abstract}
With the increasing adoption of AI-driven tools in software development, large language models (LLMs) have become essential for tasks like code generation, bug fixing, and optimization. Tools like ChatGPT, GitHub Copilot, and Codeium provide valuable assistance in solving programming challenges, yet their effectiveness remains underexplored. This paper presents a comparative study of ChatGPT, Codeium, and GitHub Copilot, evaluating their performance on LeetCode problems across varying difficulty levels and categories. Key metrics such as success rates, runtime efficiency, memory usage, and error-handling capabilities are assessed. GitHub Copilot showed superior performance on easier and medium tasks, while ChatGPT excelled in memory efficiency and debugging. Codeium, though promising, struggled with more complex problems. Despite their strengths, all tools faced challenges in handling harder problems. These insights provide a deeper understanding of each tool’s capabilities and limitations, offering guidance for developers and researchers seeking to optimize AI integration in coding workflows.
\end{abstract}


\section{Introduction}

The rise of artificial intelligence (AI) and large language models (LLMs), like GPT-4, has revolutionized software development, particularly in code generation and debugging. Trained on vast datasets, LLMs can automate complex tasks, reduce human errors, and improve programming efficiency \cite{vaswani2017attention, brown2020language, openai2023gpt}. Tools like OpenAI’s ChatGPT \cite{openai2024,openai2024chatgpt}, GitHub Copilot \cite{github2024copilot, dakhel2023github}, and Codeium \cite{codeium2024} have become popular for their abilities in code generation, real-time debugging, and problem-solving support \cite{ma2023scope, biswas2023role, gu2021domain}.

ChatGPT, built on GPT-4, has shown notable success in generating code for various domains, excelling in areas like tree algorithms but facing challenges in dynamic programming and greedy algorithms \cite{sobania2023analysis}. Similarly, GitHub Copilot, powered by Codex, automates repetitive tasks, though its performance varies across languages and environments \cite{yeticstiren2023evaluating, nygaard2024ai}. Codeium, although less extensively studied, also shows potential for boosting developer productivity \cite{kadir2024exploring}.

While AI-driven code generation tools have made notable progress, their effectiveness in competitive programming and solving complex problems remains underexplored. This study aims to address this gap by evaluating ChatGPT, Codeium, and GitHub Copilot across key metrics in competitive programming contexts. 

\section{Literature Review}

LLMs like GPT-4 have significantly influenced programming and software engineering, automating tasks such as code generation, bug fixing, and education \cite{biswas2023role, ma2023scope, anagnostopoulos2023chatgpt, sakib2023intent, hendrycks2020measuring, kasneci2023chatgpt}. 

Several studies have explored the performance of LLMs in code generation. Sobania et al. \cite{sobania2023analysis} reported ChatGPT's 71.875\% success rate on LeetCode \cite{leetcode2024}, particularly excelling in tree algorithms while struggling with dynamic programming and greedy algorithms. Prenner and Robbes \cite{prenner2021automatic} highlighted Codex's strong bug-fixing capabilities, though it hasn’t fully replaced human programmers. Yetiştiren et al. \cite{yeticstiren2023evaluating} emphasized GitHub Copilot’s effectiveness but noted inefficiencies in complex environments. 

Research on Codeium is limited, focusing mainly on its potential to enhance developer productivity without comprehensive evaluations across diverse tasks \cite{kadir2024exploring}. GitHub Copilot, extensively studied, shows limitations in handling dynamic code behaviors, reducing reliability in complex tasks \cite{ma2023scope}. Nguyen and Nadi \cite{nygaard2024ai} observed output accuracy variations depending on the programming language used, while Vaithilingam et al.\cite{vaithilingam2022expectation} identified usability challenges in aligning Copilot’s code with real-world tasks . 

In education, Biswas \cite{biswas2023role} demonstrated ChatGPT’s ability to generate and correct code in languages like C++ and Python, aiding students in numerical analysis. Kashefi and Mukerji\cite{kashefi2023chatgpt} found ChatGPT useful for debugging, though it struggled with more complex tasks . Anagnostopoulos\cite{anagnostopoulos2023chatgpt} reviewed the broader impacts of LLMs on reshaping programming education .

Empirical studies have assessed the robustness of AI tools. Hellendoorn et al. \cite{hellendoorn2019code} found that developers often prefer manual coding over using GitHub Copilot’s suggestions in challenging environments. Quoc et al.\cite{quoc2024empirical} noted the inconsistency in self-correcting models like ChatGPT . 

Despite these challenges, integrating AI tools like ChatGPT and GitHub Copilot into workflows shows promise in enhancing efficiency. However, as Salnikov noted \cite{salnikov2024artificial}, developers must remain mindful of these tools’ limitations, especially for complex tasks. Continuous refinement is essential to fully realize the potential of these AI systems \cite{sakib2023extending}.

Although several studies have examined the performance of AI tools in programming, an in-depth, comparative analysis of these tools across multiple dimensions is still needed. Our study builds on existing research by evaluating these tools within competitive programming, offering a broader perspective on their strengths and limitations.

\section{Methodology}

This section describes the systematic approach used to evaluate ChatGPT, Codeium, and GitHub Copilot across a diverse set of algorithmic challenges. We selected 300 problems from LeetCode\cite{leetcode2024}, a well-established platform known for its wide range of problems in competitive programming and technical interviews. The methodology covers dataset preparation, tool configuration, and evaluation metrics, which assess both problem-solving and debugging performance across various difficulty levels.

\subsection{Problem Selection and Dataset Preparation}

\begin{figure}[htbp]
\centering
\includegraphics[width=.6\linewidth]{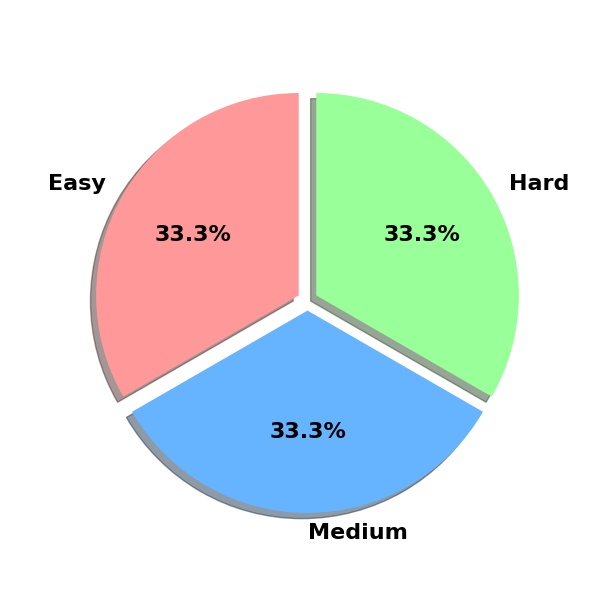}
\caption{Distribution of our dataset (300 LeetCode Problems) by Difficulty.}
\label{fig:problem_distribution}
\end{figure}

To ensure a balanced evaluation, we selected 300 LeetCode problems, equally distributed across three difficulty levels: 100 easy, 100 medium, and 100 hard. The problems were chosen to represent a broad range of algorithmic topics, including arrays, dynamic programming, graph algorithms, and recursion. This diverse set of problems is commonly used in technical interviews, making it an ideal benchmark for evaluating AI-driven programming tools. By maintaining an even distribution of problems by difficulty, we ensured that each tool was tested on challenges of varying complexity. Figure \ref{fig:problem_distribution} visually represents the distribution of problems across the three difficulty levels.

\subsection{Dataset Analysis}

\begin{figure}[htbp]
\centering
\includegraphics[width=.8\linewidth]{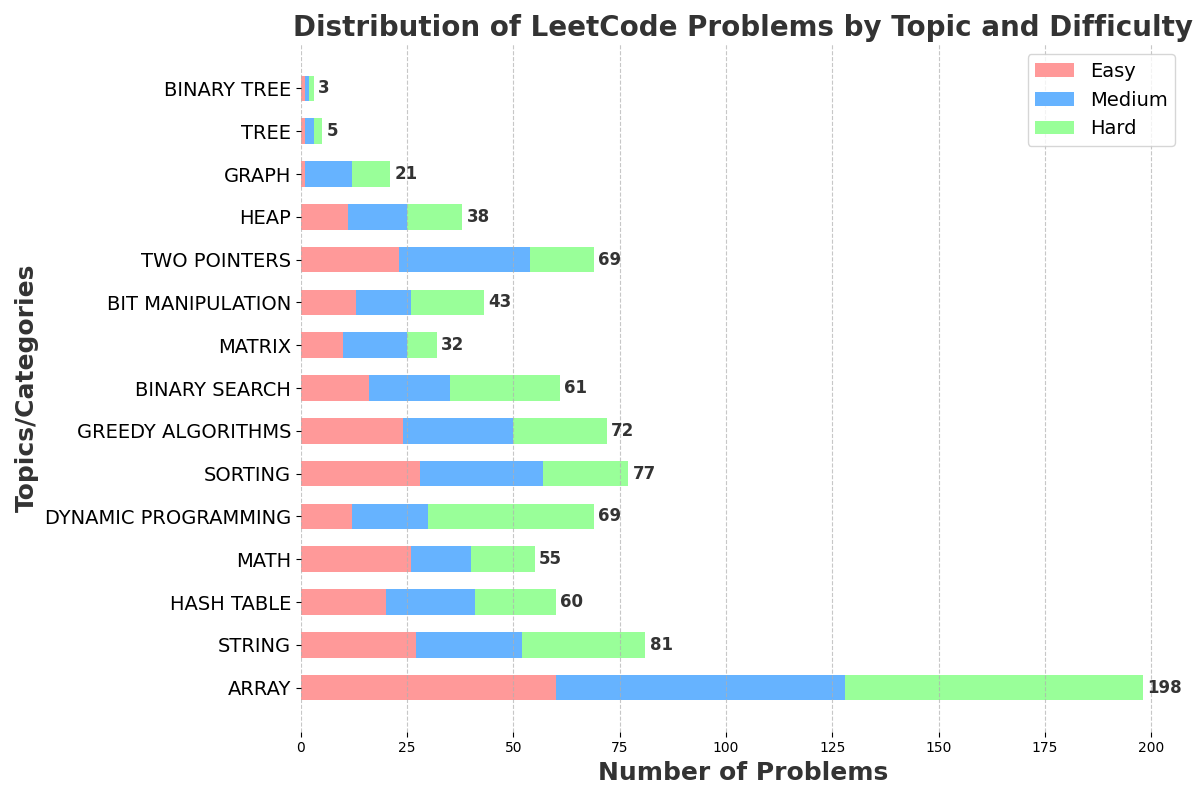}
\caption{
    Distribution of the dataset problems by topic and difficulty. The dataset is evenly distributed across difficulties (easy, medium, hard) within each topic.
}
\label{fig:topic_distribution}
\end{figure}

The selected problems span 15 distinct data structures and algorithmic topics, ensuring a comprehensive assessment of the AI tools' performance across different domains. Each problem was associated with an average of three different topics, highlighting the multi-dimensional nature of algorithmic challenges.  Figure \ref{fig:topic_distribution} provides a visual overview of the distribution of problems by topic and difficulty, showing the balanced distribution of easy, medium, and hard problems within each topic. The total number of problems for each topic is displayed at the end of each bar, covering all 15 distinct data structures and algorithmic topics.


\begin{figure}[htbp]
    \centering
    \includegraphics[width=.8\linewidth]{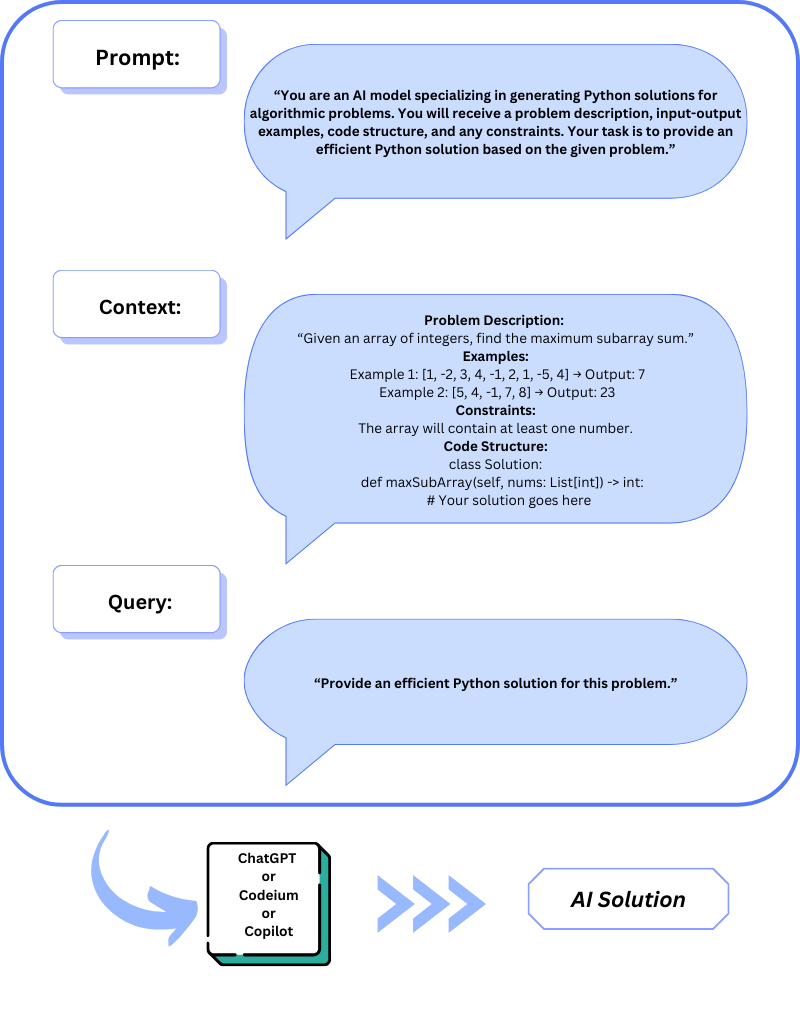}
    \caption{
        This figure illustrates the input format for the code generation task. A base prompt is passed to the AI model, which includes the problem definition, examples, constraints, and code structure. These instructions provide the necessary context for the model to understand how the task needs to be solved. Finally, a query is passed to the model, instructing it to generate the Python code solution based on the given context.
    }
    \label{fig:code_generation_task}
\end{figure}

\subsection{Tool Setup}

We configured the three AI programming assistants — ChatGPT, Codeium, and GitHub Copilot—under consistent settings for fair comparison. ChatGPT was accessed via the OpenAI API, while both Codeium and GitHub Copilot were integrated into Visual Studio Code. All tools operated with default settings, ensuring the results reflected typical user experiences without manual intervention. This uniform setup allowed us to compare the raw output of each tool directly, avoiding biases from varying configurations.

\subsection{Experimental Procedure}

The experiment was conducted in two distinct phases: \textbf{problem-solving} and \textbf{debugging}, both designed to simulate a typical software development workflow involving iterative problem-solving and debugging. Each phase aimed to test the capabilities of ChatGPT, Codeium, and GitHub Copilot in generating and correcting solutions, mimicking real-world programming tasks.

\subsubsection{Problem-Solving Phase}

In this phase, each AI tool was independently tasked with solving 300 LeetCode problems. Each tool generated solutions autonomously, without any human intervention, ensuring the results reflect the tools' inherent problem-solving capabilities. For every problem, we tracked key performance metrics such as solution accuracy, runtime efficiency, and memory usage to evaluate them.

\subsubsection{Debugging Phase}

In the debugging phase, we evaluated each tool's ability to self-correct errors. Whenever an AI tool produced an incorrect solution, it was provided with the erroneous code, error type, and detailed error information. The tool was then tasked with debugging its previous solution and generating a corrected version. This process was designed to test the tools' capacity to learn from feedback, simulating a real-world debugging scenario.

\begin{figure}[htbp]
    \centering
    \includegraphics[width=.8\linewidth]{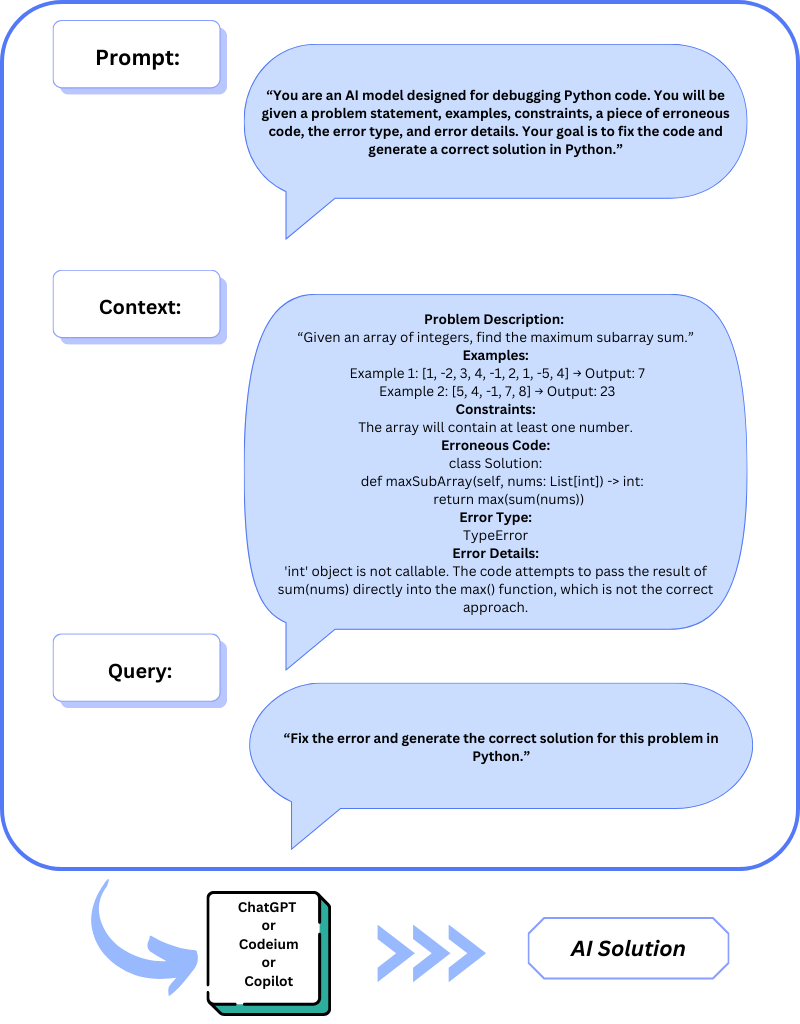}
    \caption{
        This figure presents the input structure for the debugging task, which is similar to the code generation input. The key difference lies in the prompt: for debugging, the previous erroneous code and the corresponding output are also provided. The context still includes the problem statement, examples, and constraints, but the model is asked to fix the errors and generate the correct Python solution.
    }
    \label{fig:debugging_task}
\end{figure}
\subsection{Prompt Engineering}

Prompt engineering was crucial for both the problem-solving and debugging phases of the experiment. During problem-solving, AI models were provided with structured prompts containing the problem description, examples, constraints, and required code structure. This ensured the models had the necessary context to generate accurate solutions, as shown in Figure \ref{fig:code_generation_task}.

In the debugging phase, the prompts were modified to include the incorrect code generated during the initial problem-solving phase, as well as the error type and feedback. The models were then prompted to fix the issues, testing their ability to self-correct based on the provided feedback. This approach enabled us to evaluate how well the AI models adapt to iterative workflows, similar to real-world debugging scenarios (Figure \ref{fig:debugging_task}).

\subsection{Evaluation Metrics}

The tools were evaluated using a range of metrics designed to measure both problem-solving performance and debugging efficiency:
\begin{itemize}
    \item \textbf{Success Rate:} The percentage of problems solved correctly by each tool across all difficulty levels. Submission statuses were tracked during both the problem-solving and debugging phases, including Accepted, Wrong Answer, Time Limit Exceeded, Memory Limit Exceeded, and Runtime Error. These statuses provided insight into where each tool struggled.
    \item \textbf{Runtime Efficiency:} The average runtime performance, expressed as a percentile relative to other user-submitted solutions.
    \item \textbf{Memory Efficiency:} The memory usage percentile, reflecting the solution's efficiency in terms of resource consumption.
    \item \textbf{Debugging Success Rate:} The percentage of times each tool successfully debugged its own incorrect solutions after feedback.
\end{itemize}

These metrics provide a holistic view of the tools' capabilities, including both their problem-solving efficiency and their performance in correcting errors, offering a more accurate evaluation of their overall utility for developers.

%%%%%%%%%%%%%%%%%%%%%%%%%%%%%%%%%%%%%%%%%%%%%%%%%%%%%%%%%%%%%%%%%%
%%%%%%%%%%%%%%%%%%%%%%%%%%%%%%%%%%%%%%%%%%%%%%%%%%%%%%%%%%%%%%%%%%
%%%%%%%%%%%%%%%%%%%%%%%%%%%%%%%%%%%%%%%%%%%%%%%%%%%%%%%%%%%%%%%%%%

%%%%%%%%%%%%%%%%%%%%%%%%%%%%%%%%%%%%%%%%%%%%%%%%%%%%%%%%%%%%%%%%%%
%%%%%%%%%%%%%%%%%%%%%%%%%%%%%%%%%%%%%%%%%%%%%%%%%%%%%%%%%%%%%%%%%%
%%%%%%%%%%%%%%%%%%%%%%%%%%%%%%%%%%%%%%%%%%%%%%%%%%%%%%%%%%%%%%%%%%

\section{Results and Discussion}

\subsection{Overall Performance Metrics}
\begin{figure}[htbp]
    \centering
    \includegraphics[width=\linewidth]{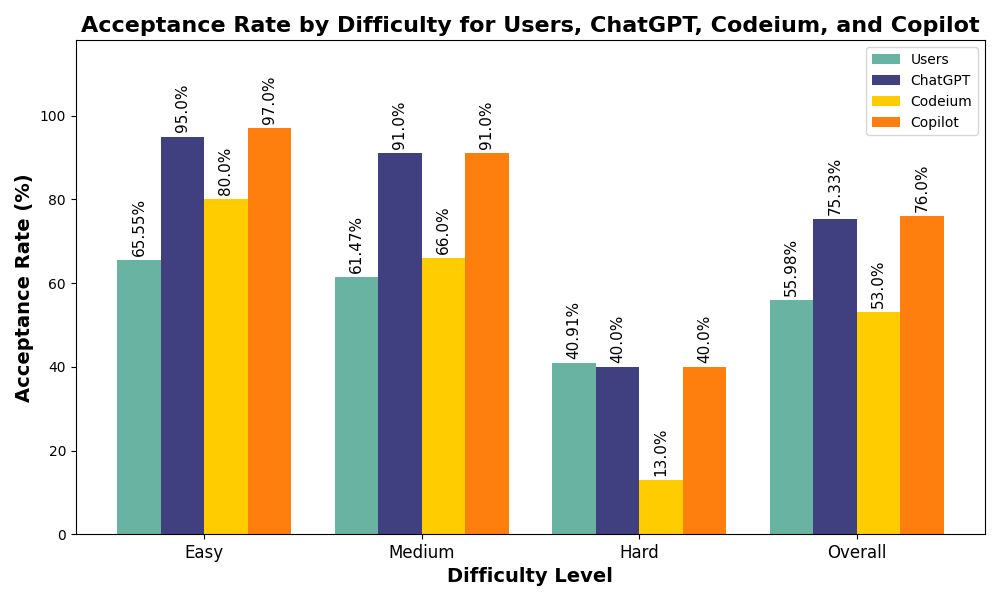}
    \caption{
        Acceptance rates for Users, ChatGPT, Codeium, and Copilot across different difficulty levels.
    }
    \label{fig:acceptance_rate}
\end{figure}

\subsubsection{Success Rate}

The success rate is a critical metric that measures how effectively each AI tool—ChatGPT, Codeium, and GitHub Copilot—solved LeetCode problems across different difficulty levels (easy, medium, and hard). Figure \ref{fig:acceptance_rate} presents an overview of acceptance rates for the tools based on difficulty, while Table 1 provides a detailed category-wise breakdown for specific problem types such as Arrays, Strings, and Hash Tables. Together, these data points offer a comprehensive view of each tool's performance across various levels of complexity and problem domains.

As depicted in Figure \ref{fig:acceptance_rate}, ChatGPT and GitHub Copilot excelled in easy and medium problems, with success rates of 95\% and 97\%, respectively, for easy problems. Both tools performed comparably on medium problems, but their performance dropped significantly on hard problems, with a success rate of 40\% for both. In comparison, users had a similar success rate of 40.91\% on hard problems, indicating that both tools struggled with the more difficult problem sets, similar to human users.

Table 1 provides more granular insights by breaking down acceptance rates for different problem categories. GitHub Copilot consistently outperformed the other tools in several categories, particularly in Arrays, where it achieved the highest overall success rate of 73.23\%, followed by ChatGPT at 71.21\%. Codeium, however, lagged behind with an overall success rate of 48.99\% in the same category. In the String category, ChatGPT led with an overall success rate of 74.07\%, closely followed by GitHub Copilot at 72.84\%. Codeium once again trailed with a lower acceptance rate of 51.85\%.

\begin{table}[htbp]
\centering
\caption{This table presents the average acceptance rate (in percentage) of solutions for LeetCode problems across different difficulty levels (Easy, Medium, Hard) and overall for Users, ChatGPT, Codeium, and GitHub Copilot. The acceptance rates reflect the proportion of problems successfully solved, illustrating each tool's performance in handling problems of varying complexity.}
\begin{tabular}{| >{\centering}m{6em} | >{\centering}m{3.2em} | >{\centering}m{2.8em} | >{\centering}m{3.4em} | >{\centering\arraybackslash}m{3.4em} | >{\centering\arraybackslash}m{3.4em} |}
  \hline
  \textbf{Category} & \textbf{Difficulty} & \textbf{Users}& \textbf{ChatGPT} & \textbf{Codeium} & \textbf{Copilot} \\
  \hline
  \hline
\multirow{4}{4em}{Array} & Easy    & 67.92  & 95.00  & 76.67  & \textbf{98.33} \\ 
\cline{2-6}
                         & Medium  & 59.77  & \textbf{89.71}  & 63.24  & 86.76 \\ 
\cline{2-6}
                         & Hard    & \textbf{39.66}  & 32.86  & 11.43  & 38.57 \\ 
\cline{2-6}
                         & Overall & 55.13  & 71.21  & 48.99  & \textbf{73.23} \\ 
\hline
  \hline
\multirow{4}{4em}{String} & Easy    & 67.41  & \textbf{96.30}  & 77.78  & 92.59 \\ 
\cline{2-6}
                          & Medium  & 63.83  & 88.00  & 72.00  & \textbf{92.00} \\ 
\cline{2-6}
                          & Hard    & 39.90  & \textbf{41.38}  & 10.34  & 37.93 \\ 
\cline{2-6}
                          & Overall & 56.46  & \textbf{74.07}  & 51.85  & 72.84 \\ 
\hline
\hline
\multirow{4}{4em}{Hash Table} & Easy    & 67.50  & \textbf{95.00}  & 80.00  & \textbf{95.00} \\ 
\cline{2-6}
                             & Medium  & 61.53  & \textbf{95.24}  & 47.62  & \textbf{95.24} \\ 
\cline{2-6}
                             & Hard    & \textbf{44.85}  & 26.32  & 31.58  & 36.84 \\ 
\cline{2-6}
                             & Overall & 58.24  & 73.33  & 53.33  & \textbf{76.67} \\ 
\hline
\hline
\multirow{4}{4em}{Math} & Easy    & 59.55  & \textbf{92.31}  & 84.62  & \textbf{92.31} \\ 
\cline{2-6}
                        & Medium  & 68.49  & 85.71  & 57.14  & \textbf{100.00} \\ 
\cline{2-6}
                        & Hard    & 41.49  & 40.00  & 0.00  & \textbf{46.67} \\ 
\cline{2-6}
                        & Overall & 56.90  & 76.36  & 54.55  & \textbf{81.82} \\ 
\hline
\hline
\multirow{4}{4em}{Dynamic Programming} & Easy    & 62.17  & \textbf{100.00}  & 91.67  & \textbf{100.00} \\ 
\cline{2-6}
                      & Medium  & 62.93  & 88.89   & 72.22  & \textbf{100.00} \\ 
\cline{2-6}
                      & Hard    & 37.86  & 35.90   & 7.69   & \textbf{43.59} \\ 
\cline{2-6}
                      & Overall & 48.63  & 60.87   & 39.13   & \textbf{68.12} \\ 
\hline
\hline
\multirow{4}{4em}{Sorting} & Easy    & 67.04  & 96.43  & 82.14  & \textbf{100.00} \\ 
\cline{2-6}
                           & Medium  & 60.72  & \textbf{93.10}  & 65.52   & 89.66 \\ 
\cline{2-6}
                           & Hard    & \textbf{47.24}  & 45.00  & 25.00   & 45.00 \\ 
\cline{2-6}
                           & Overall & 59.52  & \textbf{81.82}  & 61.04   & \textbf{81.82} \\ 
\hline
\hline
\multirow{4}{4em}{Greedy Algorithms} & Easy    & 60.19  & 83.33  & 66.67   & \textbf{91.67} \\ 
\cline{2-6}
                          & Medium  & 63.25  & \textbf{88.46}  & 65.38   & \textbf{88.46} \\ 
\cline{2-6}
                          & Hard    & \textbf{43.53}  & 40.91  & 18.18   & 36.36 \\ 
\cline{2-6}
                          & Overall & 56.21  & 72.22  & 51.39   & \textbf{73.61} \\ 
\hline
\hline
\multirow{4}{4em}{Binary Search} & Easy    & 67.18  & 93.75  & 81.25  & \textbf{100.00} \\ 
\cline{2-6}
                                 & Medium  & 54.41  & \textbf{100.00} & 63.16   & 84.21 \\ 
\cline{2-6}
                                 & Hard    & 39.54  & \textbf{42.31}  & 11.54   & 38.46 \\ 
\cline{2-6}
                                 & Overall & 51.42  & \textbf{73.77}  & 45.90   & 68.85 \\ 
\hline
\hline
\multirow{4}{4em}{Matrix} & Easy    & 76.61  & \textbf{100.00}  & 80.00  & \textbf{100.00} \\ 
\cline{2-6}
                          & Medium  & 62.35  & 80.00   & 53.33   & \textbf{86.67} \\ 
\cline{2-6}
                          & Hard    & \textbf{46.90}  & 28.57   & 0.00   & 42.86 \\ 
\cline{2-6}
                          & Overall & 63.43  & 75.00   & 50.00   & \textbf{81.25} \\ 
\hline
\hline
\multirow{4}{4em}{Bit Manipulation} & Easy    & 64.55  & \textbf{100.00}  & 84.62  & \textbf{100.00} \\ 
\cline{2-6}
                                    & Medium  & 58.63  & \textbf{100.00}  & 69.23  & \textbf{100.00} \\ 
\cline{2-6}
                                    & Hard    & 44.28  & 41.18   & 0.00   & \textbf{52.94} \\ 
\cline{2-6}
                                    & Overall & 54.75  & 76.74   & 46.51   & \textbf{81.40} \\ 
\hline
\hline
\multirow{4}{4em}{Two Pointers} & Easy    & 70.55  & \textbf{100.00}  & 91.30  & \textbf{100.00} \\ 
\cline{2-6}
                                & Medium  & 64.34  & \textbf{100.00}  & 74.19   & 93.55 \\ 
\cline{2-6}
                                & Hard    & 41.39  & \textbf{53.33}   & 0.00   & 46.67 \\ 
\cline{2-6}
                                & Overall & 61.42  & \textbf{89.86}   & 63.77   & 85.51 \\ 
\hline
\hline
\multirow{4}{4em}{Heap} & Easy    & 68.86  & \textbf{100.00}  & 81.82  & \textbf{100.00} \\ 
\cline{2-6}
                        & Medium  & 62.56  & \textbf{92.86}   & 71.43   & 78.57 \\ 
\cline{2-6}
                        & Hard    & \textbf{49.31}  & 46.15   & 23.08   & 38.46 \\ 
\cline{2-6}
                        & Overall & 59.85  & \textbf{78.95}   & 57.89   & 71.05 \\ 
\hline
\hline
\multirow{4}{4em}{Graph} & Easy    & 49.80  & \textbf{100.00}  & 0.00   & \textbf{100.00} \\ 
\cline{2-6}
                         & Medium  & 52.80  & \textbf{90.91}   & 63.64  & 81.82 \\ 
\cline{2-6}
                         & Hard    & \textbf{47.77}  & 44.44   & 22.22  & 22.22 \\ 
\cline{2-6}
                         & Overall & 50.50  & \textbf{71.43}   & 42.86  & 57.14 \\ 
\hline
\hline
\multirow{4}{4em}{Tree} & Easy    & 59.20  & \textbf{100.00}  & \textbf{100.00}  & \textbf{100.00} \\ 
\cline{2-6}
                        & Medium  & 61.80  & \textbf{100.00}  & 50.00   & \textbf{100.00} \\ 
\cline{2-6}
                        & Hard    & \textbf{52.15}  & 0.00    & 0.00    & 0.00 \\ 
\cline{2-6}
                        & Overall & 57.42  & \textbf{60.00}   & 40.00   & \textbf{60.00} \\ 
\hline
\hline
\multirow{4}{4em}{Binary Tree} & Easy    & 59.20  & \textbf{100.00}  & \textbf{100.00}  & \textbf{100.00} \\ 
\cline{2-6}
                               & Medium  & 57.90  & \textbf{100.00}  & \textbf{100.00}  & \textbf{100.00} \\ 
\cline{2-6}
                               & Hard    & \textbf{37.50}  & 0.00    & 0.00    & 0.00 \\ 
\cline{2-6}
                               & Overall & 51.53  & \textbf{66.67}   & \textbf{66.67}   & \textbf{66.67} \\ 
\hline
\end{tabular}
\end{table}

Despite strong performances on easier problems, Codeium struggled significantly with hard problems across all categories, with an overall success rate as low as 13\% for hard problems. This trend is particularly evident in categories like Arrays and Sorting, where Codeium’s performance on hard problems was notably weaker than both ChatGPT and GitHub Copilot. Users, meanwhile, consistently performed well on hard problems, with higher success rates in categories such as Arrays (39.66\%) and Sorting (47.24\%), compared to the AI tools.

In summary, GitHub Copilot demonstrated the best overall performance, particularly on easy and medium problems, with slightly better success rates than ChatGPT across most categories. However, both tools faced challenges with hard problems, aligning their performance more closely with that of human users. Codeium, while effective on easier problems, struggled considerably with harder problem sets, particularly in complex categories such as Sorting and Dynamic Programming.

\begin{figure}[htbp]
    \centering
    \includegraphics[width=.9\linewidth]{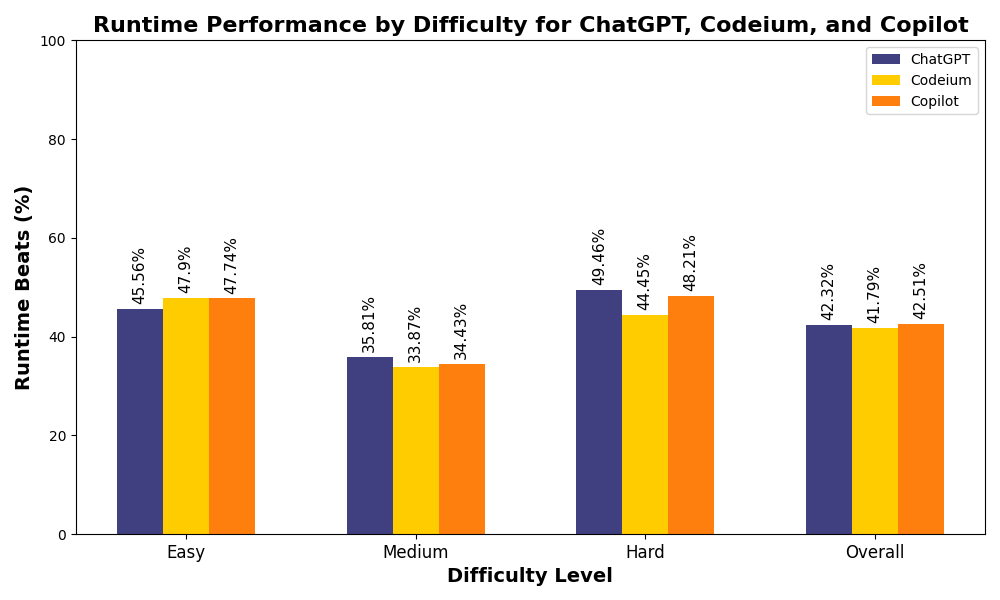}
    \caption{
    Bar chart showing the runtime performance for ChatGPT, Codeium, and Copilot across different difficulty levels.
    }
    \label{fig:runtime_performance}
\end{figure}
\begin{figure}[htbp]
    \centering
    \includegraphics[width=.9\linewidth]{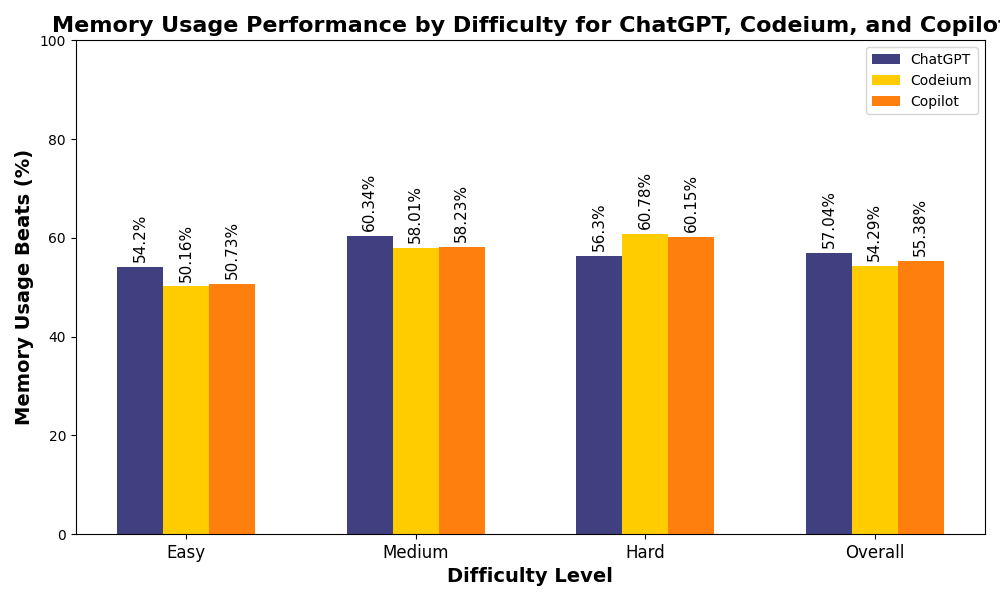}
    \caption{
    Bar chart representing memory usage efficiency for ChatGPT, Codeium, and Copilot.
    }
    \label{fig:memory_usage_performance}
\end{figure}

\subsubsection{Runtime Performance}

The runtime performance of each tool was evaluated based on how their solution execution times compared to user-submitted solutions, as shown in Figure \ref{fig:runtime_performance}. All three tools—ChatGPT, Codeium, and GitHub Copilot—exhibited comparable runtime efficiency for easy problems, with Copilot slightly outperforming the others. For medium and hard problems, ChatGPT showed stronger performance, particularly for hard problems, though the differences between the tools were minor. Overall, Copilot held a slight advantage across difficulty levels, indicating marginally better runtime efficiency overall.

\subsubsection{Memory Usage}

Memory usage efficiency was assessed by comparing the tools' memory consumption to other user-submitted solutions, as depicted in Figure \ref{fig:memory_usage_performance}. ChatGPT consistently performed best in terms of memory usage, especially for medium problems. Codeium and Copilot, while close in performance, showed slightly higher memory usage on hard problems compared to ChatGPT. Overall, ChatGPT proved to be the most memory-efficient tool across all difficulty levels, making it the strongest performer in this category.

\subsection{Error Handling and Debugging}

In the debugging phase, each tool was tasked with fixing its own incorrect solutions based on the feedback provided. After an error in the initial problem-solving phase, the tool was supplied with the erroneous code, error type (e.g., Wrong Answer, Runtime Error), and detailed feedback, but not the correct solution. This tested the tools’ ability to learn from mistakes and simulate real-world debugging.

\begin{figure}[htbp]
    \centering
    \includegraphics[width=\linewidth]{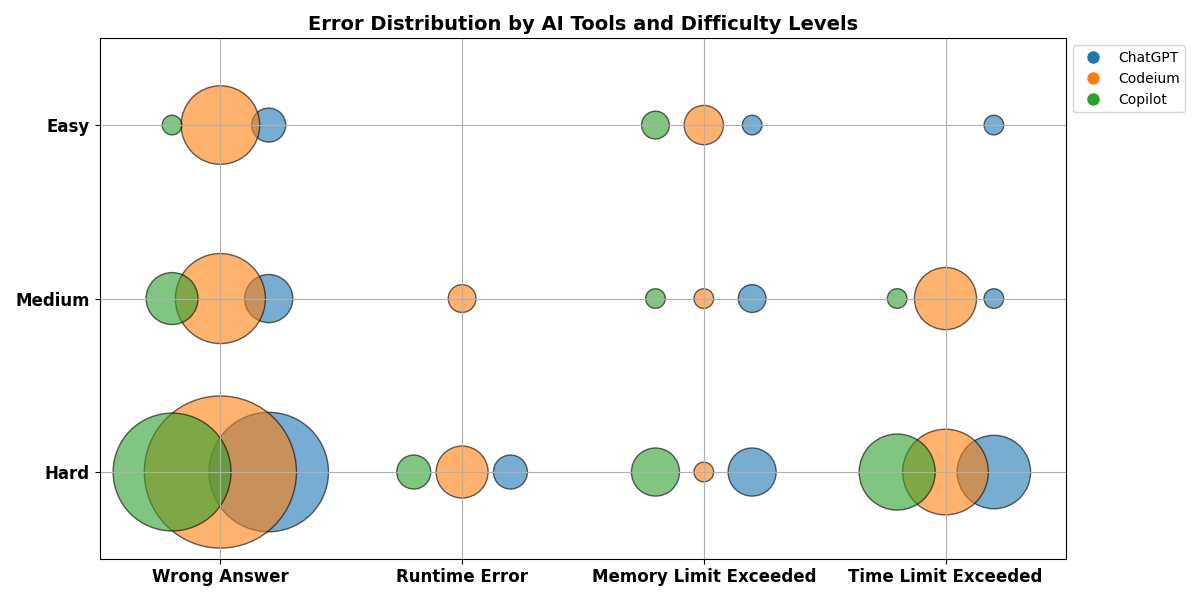}
    \caption{
        Error distribution for AI tools across three difficulty levels. The figure shows the percentage of four types of errors represented by the size of the bubbles for each AI tool.
    }
    \label{fig:bubble_chart_errors}
\end{figure}

Figure \ref{fig:bubble_chart_errors} shows the error distribution across difficulty levels, where Codeium encountered the most frequent errors, especially on hard problems, while GitHub Copilot and ChatGPT had fewer issues. Wrong Answer and Time Limit Exceeded errors were particularly high for Codeium in challenging problems.

\begin{table}[htbp]
\centering
\caption{Debugging Success Rate for ChatGPT, Codeium, and GitHub Copilot across problem difficulties.}
\begin{tabular}{|c|c|c|c|}
\hline
\textbf{Difficulty Level} & \textbf{ChatGPT} & \textbf{Codeium} & \textbf{GitHub Copilot} \\
\hline
Easy & 100\% & 80\% & 100\% \\
\hline
Medium & 66.67\% & 52.94\% & 55.56\% \\
\hline
Hard & 42.5\% & 20.69\% & 40\% \\
\hline
\end{tabular}
\label{tab:debugging}
\end{table}

Table \ref{tab:debugging} outlines the debugging success rates. ChatGPT demonstrated the best performance overall, with a success rate of 42.5\% on hard problems, followed by GitHub Copilot at 40\%. Codeium struggled significantly, correcting only 20.69\% of its errors in hard problems. These results highlight ChatGPT’s superior self-correction capabilities, especially in more complex scenarios, while GitHub Copilot and Codeium faced more challenges in adapting to difficult problem sets.

\section{Threats to Validity}

This study presents several potential limitations. The dataset comprised 300 LeetCode problems, focused on key areas of data structures and algorithms. While balanced across difficulty levels, it does not encompass all algorithmic domains, and a larger, more diverse set could provide deeper insights. Additionally, the results are based on LeetCode submission statistics, which may not fully generalize to other platforms like Codeforces, HackerRank, or CodeChef, where problem styles and difficulty may vary. Moreover, the AI models—ChatGPT, Codeium, and GitHub Copilot—are continuously evolving. Our evaluation, conducted in mid-2024, may not reflect future performance as these tools incorporate new data and improve. Researchers replicating or extending this study may observe different outcomes due to these updates.

\section{Conclusion}

A detailed comparison of ChatGPT, Codeium, and GitHub Copilot reveals key insights into their strengths and limitations across success rate, runtime efficiency, memory usage, and error handling. GitHub Copilot consistently demonstrated the highest success rate, excelling in easy and medium tasks. However, both Copilot and ChatGPT struggled with hard problems, achieving a 40\% success rate, comparable to human users. ChatGPT proved the most efficient in terms of memory usage, especially for medium problems, while GitHub Copilot exhibited slightly better runtime performance overall. 

In the debugging phase, ChatGPT emerged as the most effective, successfully correcting 42.5\% of errors on hard problems, demonstrating a strong ability to learn from feedback. Codeium, while performing well on easier tasks, lagged behind on harder problems, both in problem-solving and debugging. These findings underscore the strengths and limitations of each tool, showing that while they are highly effective in certain areas, none are yet capable of consistently outperforming human problem-solving abilities in more complex scenarios.

%%%%%%%%%%%%%%%%%%%%%%%%%%%%%%%%%%%%%%%%%%%%%%%%%%%%%%%%%%%%%%%%%
%%%%%%%%%%%%%%%%%%%%%%%%%%%%%%%%%%%%%%%%%%%%%%%%%%%%%%%%%%%%%%%%%
%%%%%%%%%%%%%%%%%%%%%%%%%%%%%%%%%%%%%%%%%%%%%%%%%%%%%%%%%%%%%%%%%



\begin{thebibliography}{10}
\providecommand{\url}[1]{#1}
\csname url@samestyle\endcsname
\providecommand{\newblock}{\relax}
\providecommand{\bibinfo}[2]{#2}
\providecommand{\BIBentrySTDinterwordspacing}{\spaceskip=0pt\relax}
\providecommand{\BIBentryALTinterwordstretchfactor}{4}
\providecommand{\BIBentryALTinterwordspacing}{\spaceskip=\fontdimen2\font plus
\BIBentryALTinterwordstretchfactor\fontdimen3\font minus \fontdimen4\font\relax}
\providecommand{\BIBforeignlanguage}[2]{{%
\expandafter\ifx\csname l@#1\endcsname\relax
\typeout{** WARNING: IEEEtran.bst: No hyphenation pattern has been}%
\typeout{** loaded for the language `#1'. Using the pattern for}%
\typeout{** the default language instead.}%
\else
\language=\csname l@#1\endcsname
\fi
#2}}
\providecommand{\BIBdecl}{\relax}
\BIBdecl

\bibitem{vaswani2017attention}
A.~Vaswani, ``Attention is all you need,'' \emph{Advances in Neural Information Processing Systems}, 2017.

\bibitem{brown2020language}
T.~B. Brown, ``Language models are few-shot learners,'' \emph{arXiv preprint arXiv:2005.14165}, 2020.

\bibitem{openai2023gpt}
R.~OpenAI, ``Gpt-4 technical report. arxiv 2303.08774,'' \emph{View in Article}, vol.~2, no.~5, 2023.

\bibitem{openai2024}
\BIBentryALTinterwordspacing
OpenAI, ``Openai,'' 2024, [Accessed September 2024]. [Online]. Available: \url{https://openai.com}
\BIBentrySTDinterwordspacing

\bibitem{openai2024chatgpt}
\BIBentryALTinterwordspacing
------, ``Chatgpt,'' 2024, [Accessed September 2024]. [Online]. Available: \url{https://openai.com/chatgpt}
\BIBentrySTDinterwordspacing

\bibitem{github2024copilot}
\BIBentryALTinterwordspacing
GitHub, ``Github copilot,'' 2024, [Accessed September 2024]. [Online]. Available: \url{https://github.com/features/copilot}
\BIBentrySTDinterwordspacing

\bibitem{dakhel2023github}
A.~M. Dakhel, V.~Majdinasab, A.~Nikanjam, F.~Khomh, M.~C. Desmarais, and Z.~M.~J. Jiang, ``Github copilot ai pair programmer: Asset or liability?'' \emph{Journal of Systems and Software}, vol. 203, p. 111734, 2023.

\bibitem{codeium2024}
\BIBentryALTinterwordspacing
Codeium, ``Codeium: Ai-powered code autocomplete,'' 2024, [Accessed September 2024]. [Online]. Available: \url{https://codeium.com}
\BIBentrySTDinterwordspacing

\bibitem{ma2023scope}
W.~Ma, S.~Liu, W.~Wang, Q.~Hu, Y.~Liu, C.~Zhang, L.~Nie, and Y.~Liu, ``The scope of chatgpt in software engineering: A thorough investigation,'' \emph{arXiv preprint arXiv:2305.12138}, 2023.

\bibitem{biswas2023role}
S.~Biswas, ``Role of chatgpt in computer programming.'' \emph{Mesopotamian Journal of Computer Science}, vol. 2023, pp. 9--15, 2023.

\bibitem{gu2021domain}
Y.~Gu, R.~Tinn, H.~Cheng, M.~Lucas, N.~Usuyama, X.~Liu, T.~Naumann, J.~Gao, and H.~Poon, ``Domain-specific language model pretraining for biomedical natural language processing,'' \emph{ACM Transactions on Computing for Healthcare (HEALTH)}, vol.~3, no.~1, pp. 1--23, 2021.

\bibitem{sobania2023analysis}
D.~Sobania, M.~Briesch, C.~Hanna, and J.~Petke, ``An analysis of the automatic bug fixing performance of chatgpt,'' in \emph{2023 IEEE/ACM International Workshop on Automated Program Repair (APR)}.\hskip 1em plus 0.5em minus 0.4em\relax IEEE, 2023, pp. 23--30.

\bibitem{yeticstiren2023evaluating}
B.~Yeti{\c{s}}tiren, I.~{\"O}zsoy, M.~Ayerdem, and E.~T{\"u}z{\"u}n, ``Evaluating the code quality of ai-assisted code generation tools: An empirical study on github copilot, amazon codewhisperer, and chatgpt,'' \emph{arXiv preprint arXiv:2304.10778}, 2023.

\bibitem{nygaard2024ai}
J.~Nyg{\aa}rd, ``Ai-assisted code generation tools,'' Master's thesis, J. Nyg{\aa}rd, 2024.

\bibitem{kadir2024exploring}
M.~E. Kadir, T.~Rahman, S.~Barman, and M.~Al-Amin, ``Exploring the competency of chatgpt in solving competitive programming challenges,'' \emph{International Journal}, vol.~13, no.~1, 2024.

\bibitem{anagnostopoulos2023chatgpt}
C.-N. Anagnostopoulos, ``Chatgpt impacts in programming education: A recent literature overview that debates chatgtp responses,'' \emph{arXiv preprint arXiv:2309.12348}, 2023.

\bibitem{sakib2023intent}
F.~A. Sakib, A.~Karim, S.~H. Khan, and M.~M. Rahman, ``Intent detection and slot filling for home assistants: Dataset and analysis for bangla and sylheti,'' \emph{arXiv preprint arXiv:2310.10935}, 2023.

\bibitem{hendrycks2020measuring}
D.~Hendrycks, C.~Burns, S.~Basart, A.~Zou, M.~Mazeika, D.~Song, and J.~Steinhardt, ``Measuring massive multitask language understanding,'' \emph{arXiv preprint arXiv:2009.03300}, 2020.

\bibitem{kasneci2023chatgpt}
E.~Kasneci, K.~Se{\ss}ler, S.~K{\"u}chemann, M.~Bannert, D.~Dementieva, F.~Fischer, U.~Gasser, G.~Groh, S.~G{\"u}nnemann, E.~H{\"u}llermeier \emph{et~al.}, ``Chatgpt for good? on opportunities and challenges of large language models for education,'' \emph{Learning and individual differences}, vol. 103, p. 102274, 2023.

\bibitem{leetcode2024}
\BIBentryALTinterwordspacing
LeetCode, ``Leetcode: Online judge for algorithm problems,'' 2024, [Accessed September 2024]. [Online]. Available: \url{https://leetcode.com}
\BIBentrySTDinterwordspacing

\bibitem{prenner2021automatic}
J.~A. Prenner and R.~Robbes, ``Automatic program repair with openai's codex: Evaluating quixbugs,'' \emph{arXiv preprint arXiv:2111.03922}, 2021.

\bibitem{vaithilingam2022expectation}
P.~Vaithilingam, T.~Zhang, and E.~L. Glassman, ``Expectation vs. experience: Evaluating the usability of code generation tools powered by large language models,'' in \emph{Chi conference on human factors in computing systems extended abstracts}, 2022, pp. 1--7.

\bibitem{kashefi2023chatgpt}
A.~Kashefi and T.~Mukerji, ``Chatgpt for programming numerical methods,'' \emph{Journal of Machine Learning for Modeling and Computing}, vol.~4, no.~2, 2023.

\bibitem{hellendoorn2019code}
V.~J. Hellendoorn, S.~Proksch, H.~C. Gall, and A.~Bacchelli, ``When code completion fails: A case study on real-world completions,'' in \emph{2019 IEEE/ACM 41st International Conference on Software Engineering (ICSE)}.\hskip 1em plus 0.5em minus 0.4em\relax IEEE, 2019, pp. 960--970.

\bibitem{quoc2024empirical}
T.~T. Quoc, D.~H. Minh, T.~Q. Thanh, and A.~Nguyen-Duc, ``An empirical study on self-correcting large language models for data science code generation,'' \emph{arXiv preprint arXiv:2408.15658}, 2024.

\bibitem{salnikov2024artificial}
D.~Salnikov, ``Artificial intelligence in software engineering,'' 2024.

\bibitem{sakib2023extending}
F.~A. Sakib, S.~H. Khan, and A.~Karim, ``Extending the frontier of chatgpt: Code generation and debugging,'' \emph{arXiv preprint arXiv:2307.08260}, 2023.

\end{thebibliography}
\end{document}